\documentclass{aa}  

\usepackage{graphicx}
\usepackage[dvipsnames]{xcolor}
\graphicspath{ {./figures/} }
\usepackage{txfonts}
\usepackage[%
  breaklinks = true,
  colorlinks = true,
  urlcolor   = blue,
  citecolor  = blue,
  linkcolor  = blue,
]{hyperref}
\usepackage{soul}
\usepackage[labelfont=bf]{caption}

\newcommand{\yokshi}{YS-94}
\newcommand{\syntelis}{Syntelis-19}
\newcommand{\hdiff}{Gudiksen-11}

\begin{document}

   \title{A comparative study of resistivity models for simulations of magnetic reconnection in the solar atmosphere}

   \titlerunning{Comparative study of resistivity models}
   \authorrunning{Ø.H.Færder et al.}

   \author{Ø. H. Færder\inst{1,2},
          D. Nóbrega-Siverio\inst{3,4,1,2}
          \and
          M. Carlsson\inst{1,2}
          }

   \institute{Rosseland Centre of Solar Physics, University of Oslo,
              PO Box 1029, Blindern, NO-0315 Oslo, Norway\\
              \email{o.h.farder@astro.uio.no}
         \and
              Institute of Theoretical Astrophysics, University of Oslo,
              PO Box 1029, Blindern, NO-0315 Oslo, Norway
         \and
             Instituto de Astrof\'isica de Canarias,    E-38205 La Laguna,  Tenerife, Spain
        \and
            Universidad de La Laguna, Dept. Astrof\'isica, E-38206 La  Laguna, Tenerife, Spain
             }

  \abstract
  {Magnetic reconnection is a fundamental mechanism in astrophysics. A common challenge in mimicking this process numerically in particular for the Sun is that the solar electrical resistivity is small compared to the diffusive effects caused by the discrete nature of codes.}
   {We aim to study different anomalous resistivity models and their respective effects on  simulations related to magnetic reconnection in the Sun.}
    {We used the Bifrost code to perform a 2D numerical reconnection experiment in the corona that is driven by converging opposite polarities at the solar surface. This experiment was run with three different commonly used resistivity models: 1) the hyper-diffusion model originally implemented in Bifrost, 2) a resistivity proportional to the current density, and 3) a resistivity proportional to the square of the electron drift velocity. The study was complemented with a 1D experiment of a Harris current sheet with the same resistivity models.}
   {The 2D experiment shows that the three resistivity models are capable of producing results in satisfactory agreement with each other in terms of the current sheet length, inflow velocity, and Poynting influx. Even though Petschek-like reconnection occurred with the current density-proportional resistivity while the other two cases mainly followed plasmoid-mediated reconnection,  the large-scale evolution of thermodynamical quantities such as temperature and density are quite similar between the three cases. For the 1D experiment, some recalibration of the diffusion parameters is needed to obtain comparable results. Specifically the hyper-diffusion and the drift velocity-dependent resistivity model needed only minor adjustments, while the current density-proportional model needed a rescaling of several orders of magnitude.}
  {The Bifrost hyper-diffusion model is as suitable for simulations of magnetic reconnection as other common resistivity models and has the advantage of being applicable to any region in the solar atmosphere without the need for significant recalibration.}

   \keywords{
                magnetohydrodynamics (MHD) --
                magnetic reconnection --
                methods: numerical -- Sun: atmosphere -- Sun: corona -- Sun: magnetic fields
               }

   \maketitle
   
%

\section{Introduction}

Magnetic reconnection plays a crucial role in a wide range of phenomena in the Universe. 
For instance, it sparks high-energetic bursts in the accretion disc around the black hole in active galactic nuclei \citep{2002ApJ...572L.173L}, it is the basis of thermonuclear power devices, such as the tokamak \citep{1973PhFl...16.1054F}, and it strongly affects space weather \citep{1979Natur.282..243P}. On the Sun in particular, this physical process has been shown through numerical experiments to cause several remarkable solar events, such as 
Ellerman bombs (EBs) and ultraviolet (UV) bursts \citep[e.g.][]{               2017ApJ...839...22H, 
    2019A&A...626A..33H,
    2017A&A...601A.122D,
    2017ApJ...850..153N,
    2019A&A...628A...8P,
    2021A&A...646A..88N},
surges and coronal jets \citep[e.g.][]{   
    1995Natur.375...42Y, 1996PASJ...48..353Y,
    2008ApJ...683L..83N,
    2009ApJ...691...61P,
    2013ApJ...771...20M,
    2013ApJ...769L..21A,
    2014ApJ...789L..19F,
    2015ApJ...811..138T,
    2016ApJ...822...18N, 
    2016ApJ...827....4W,
    2017Natur.544..452W,
    2017ApJ...834...62K,
    2021ApJ...912...75L,
    2022ApJ...935L..21N},
and flares \citep[e.g.][]{
    2001ApJ...549.1160Y,
    2009ApJ...700..559M,
    2019NatAs...3..160C,
    2023arXiv230305299R,
    2023arXiv230305405C},
to mention some.

Theoretical reconnection models are commonly divided into two types: slow-reconnection and fast-reconnection. The slow-reconnection model developed by  
\citet{1958IAUS....6..123S, 1958NCim....8S.188S} and \citet{1957JGR....62..509P} assumes constant diffusivity over the whole reconnection site and predicts exactly one-half of the inflowing magnetic energy to be converted into heat and the other half into kinetic energy. Nonetheless, the Sweet-Parker model is not efficient enough to reproduce the relatively high reconnection rate observed in flares \citep[e.g. ][and references therein]{2014masu.book.....P}. The fast-reconnection model developed by \cite{1964NASSP..50..425P}  instead assumes a diffusion layer limited to a small segment of the boundary layer between the opposing magnetic fields with 
slow-mode shock waves
propagating from the diffusion region. Most of the energy conversion in this model takes place at the shocks, and for a specific heat ratio of $\gamma = \frac{5}{3}$, two-fifths of the inflowing magnetic energy is turned into heat and the remaining three-fifths into kinetic energy. This model predicts a reconnection rate that is high enough to reproduce flares. The Sweet-Parker model and the Petschek model are both steady-state models that assume that the current sheets are stable and do not break. However, reconnection theory has shown that current sheets tend to undergo different resistive instabilities, such as 
the tearing instability\citep{1963PhFl....6..459F}, causing plasmoids (magnetic islands) to appear and move along the current-flow lines.
As a consequence, the reconnection rate and energy conversion rate may deviate from the values predicted analytically with the Sweet-Parker and the Petschek model, and careful analysis is therefore required when studying non-stationary reconnection through numerical simulations.

Mimicking magnetic reconnection processes from a numerical perspective is challenging due to the complex behaviour of the electrical resistivity, $\eta$, which appears in Ohm's law as the ratio of the electric field strength and the current density in the rest frame of the fluid. In the solar atmosphere, this coefficient is commonly derived from kinetic theory of particle collisions and given by Spitzer resistivity \citep{1962pfig.book.....S}. However, under some conditions, such as regions of strong magnetic field gradients, plasma instabilities can affect the dynamics of the charged particles and can cause the resistivity to rise beyond the Spitzer value \citep{2002A&A...382..639R}. This effect, known as anomalous resistivity, is also a necessary component to support the theory of dissipation of direct currents \citep{1984A&A...137...63H} as a significant source of coronal heating because the collisional Spitzer resistivity is too small to dissipate such strong currents \citep{2013A&A...557A.118A}. In addition, we need to take into account the diffusive effects caused by the discrete nature of numerical codes, which are often significantly greater than those caused by the physical resistivity. Especially in numerical models of the solar atmosphere, regions of large magnetic field gradients require a diffusivity that is much larger than the Spitzer resistivity in order to become numerically resolvable.
Because of this, it is common to apply ad hoc terms
for anomalous resistivity \citep{1979PhFl...22.1189S,1995NordlundGalsgaard, 2002A&A...382..639R, 2005A&A...429..335V, 2010ApJ...719..357F, 2013A&A...557A.118A, 2014ApJ...789..132R, 2017ApJ...834...10R, 2022A&A...664A..91P} that are set to be large around current sheets in order to dissipate them until they become numerically resolvable, but stay small elsewhere in order to keep the Reynolds and Lundquist numbers relatively high. 

For a steady Sweet-Parker- or Petschek-like reconnection model, it is sufficient to use a localised anomalous resistivity model, which means that the resistivity is set to a non-zero value (or to a function of spatial coordinates) in a specific location and zero elsewhere \citep{1999SoPh..185..127I}. Non-steady reconnection models with a plasmoid instability can be simulated by using a more adaptive anomalous resistivity model, for instance by enhancing the resistivity when the electron drift velocity or the current density surpass a given threshold value \citep[e.g.][]{1979PhFl...22.1189S}, or by applying a fourth-order hyper-diffusive operator consisting of a small global diffusive term and a location-specific diffusion term \citep[e.g.][]{1995NordlundGalsgaard, 2011A&A...531A.154G}. However, if the numerical resolution is sufficiently high in areas of strong magnetic field gradients, it is even possible to successfully simulate reconnection with a plasmoid instability without adding any anomalous resistivity terms and only using the actual resistivity in the solar atmosphere \citep[e.g.][]{2021A&A...646A..88N}.

In this paper, three different resistivity models are applied on two numerical experiments for the purpose of analysing their effects on magnetic reconnection. The first experiment mimics a 2D simulation by \citet{2019ApJ...872...32S}. This enables us to compare our results with already published results that were obtained using a different numerical code. The second experiment simulates a 1D Harris current sheet. We can therefore study the diffusive effects that the resistivity models have in a simple setup.

The structure of the paper is as follows. Section~\ref{sec:model} describes the numerical code and the model equations (Sect.~\ref{sec:model-equations}) we used for our experiments, the resistivity models (Sect.~\ref{sec:resistivity-models}), and the setup for the numerical experiments (Sect.~\ref{sec:experiments}). Section~\ref{sec:Results} gives a detailed analysis of the results for the 2D experiment (Sect.~\ref{sec:Results-2Dflux}) and the 1D experiment (Sect.~\ref{sec:Results-1Dharris}). Finally,  Sect.~\ref{sec:discussion} contains a brief discussion of the key results of our study and summarises the conclusions.

%
%
\section{Numerical model}
\label{sec:model}
The simulations of this paper were performed with the Bifrost code \citep{2011A&A...531A.154G}. Bifrost is a massively parallel 3D code that solves the
equations of magnetohydrodynamics (MHD)
on a staggered grid using a sixth-order differential operator to discretise the spatial derivatives, supported by fifth-order interpolation operators. For the time-stepping, we chose a third-order method \citep{1979acmp.proc..313H}. The code is modular and can take various physical ingredients into account depending on the experiment.

\subsection{Model equations}
\label{sec:model-equations}

The model equations for our experiments are given by

\begin{align}
    \frac{\partial \rho}{\partial t} &=-\nabla\cdot (\rho \mathbf{u}) , \label{eq:continuity} \\
    \frac{\partial (\rho\mathbf{u})}{\partial t} &=-\nabla\cdot \left( \rho\mathbf{u}\otimes\mathbf{u} - \bar{\bar{\tau}}  \right) - \nabla P + \mathbf{J}\times\mathbf{B} + \rho\mathbf{g} , \label{eq:momentum} \\
    \frac{\partial\mathbf{B}}{\partial t} &= -\nabla\times ( - \mathbf{u}\times\mathbf{B} + \bar{\bar{\eta}}\mathbf{J}) \label{eq:induction} ,  \\
    \frac{\partial e}{\partial t} &= -\nabla\cdot (e\mathbf{u}) - P \nabla\cdot\mathbf{u} + Q_J + Q_V + Q_C , 
    \label{eq:energy}
\end{align}
\noindent
where $\rho$, $\mathbf{u}$, $e$, and $\mathbf{B}$ are the mass density, fluid velocity, internal energy per unit volume, and the magnetic field, respectively. 
$\bar{\bar{\tau}}$, $P$, $\mathbf{J}$, $\mathbf{g}$, $\bar{\bar{\eta}}$, $Q_J$, $Q_V$, and $Q_C$ are the viscous stress tensor, gas pressure, electric current density, gravitational acceleration, electrical resistivity tensor, Joule heating, viscous heating, and the Spitzer thermal conductivity term, respectively. Other terms  such as non-equilibrium ionisation, ambipolar diffusion, Hall effect, radiative cooling, and optically thin losses are neglected in our experiments. The gravitational term $\rho \mathbf{g}$, with $g=0.274\ \mathrm{km\ s^{-2}}$, and the Spitzer thermal conductivity term $Q_C$ are only included in the first experiment of this paper (Sect.~\ref{sec:syntelis}).

For the equation-of-state, we used the same equation as \citet{2019ApJ...872...32S}, that is, an electrically neutral ideal gas with a specific heat ratio of $\gamma=\frac{5}{3}$ and a mean molecular weight of $\mu=1.2$, where $P$ and $e$ are related to the mass density, $\rho$, and temperature, $T$, as follows:
\begin{align}
    P &= \frac{\rho k_B T}{\mu m_H} ,\label{eq:ideal_gas-P} \\
    e &= \frac{P}{(\gamma - 1)} , \label{eq:ideal_gas-e} 
\end{align}
where $k_B$ and $m_H$ are the Boltzmann constant and mass of hydrogen, respectively.

\subsection{Electrical resistivity models}
\label{sec:resistivity-models}

For the purpose of analysing the effects of the electrical resistivity model on the reconnection in the corona, three different approaches were compared:
1) the default way of handling magnetic resistivity in Bifrost, by means of hyper-diffusion \citep{2011A&A...531A.154G}, hereafter referred to as the {\hdiff} model (see Sect. \ref{sec:hyper-model}),
2) a resistivity that scales linearly with the current density as was used by \cite{2019ApJ...872...32S} for their 2D flux cancellation simulation, which is mimicked in this paper (see Sect. \ref{sec:syntelis}),
hereafter referred to as the {\syntelis} model (see Sect. \ref{sec:s19-model});
 and 3) a resistivity that scales quadratically with the electron drift velocity employed by \cite{1994ApJ...436L.197Y} for their simulation of an emerging coronal loop, hereafter referred to as the {\yokshi} model. Inspired by \cite{1979PhFl...22.1189S}, the latter resistivity model
has been used in several other papers \citep[e.g.][]{1992PASJ...44..265S,1993ASPC...46..500S,1996PASJ...48..353Y,2004PThPS.155..124M}.

For later reference, we introduce here the definitions of the Reynolds number, $Re$, and Lundquist number, $S_L$,
\begin{align}
    Re \equiv  \frac{|{\bf u}| L_B}{\eta}, \label{eq:reynolds} \\
    S_L \equiv \frac{v_A L_B}{\eta}, \label{eq:lundquist}
\end{align}
where $L_B \equiv (|{\bf J}|/|{\bf B}|)^{-1}$ is the characteristic length of the magnetic field, and $v_A \equiv |{\bf B}|/\sqrt{\mu_0 \rho}$ is the Alfvén speed of the plasma, where $\mu_0$ is the vacuum permeability.

\subsubsection{{\hdiff} model}\label{sec:hyper-model}
Based on the resistivity model developed by \citet{1995NordlundGalsgaard}, 
the {\hdiff} resistivity consists of two major terms.
The first term is an electrical diffusive speed, ${\bf U_m}$, with the $x_i$ component  defined by 
\begin{align}
    U_{m,i} =  \nu_1 c_f + \nu_2 |u_i| + \eta_3 \Delta x_i |\nabla_{\perp} u_i| , \label{eq:diffusive_velocity_Um}
\end{align}
where $\nu_1$, $\nu_2$, and $\eta_3$ are scaling factors for the fast-mode wave velocity, bulk velocity, and gradients in the velocity perpendicular to the magnetic field, respectively; and $c_f \equiv \sqrt{c_s^2 + v_A^2}$ is the fast-mode speed, with the sound speed $c_s$ given by $c_s \equiv \sqrt{\gamma P/\rho}$. In our experiments, we set $\nu_1 = 0.03, \nu_2 = 0.2,$ and $\eta_3 = 0.2$, which are typical values used in Bifrost simulations. In Sect.~\ref{sec:results-2dflux-freeparameters} we discuss how modifying these free parameters affects the results.

The second term is a positive definite quenching operator defined by
\begin{align}
    \mathbb{Q}_i(g) \equiv \frac{|\Delta_i^2 g|}{|g| + |\Delta_i^2 g|/q_{max}} ,
\end{align}
where $\Delta_i^2$ is the second-order difference operator in the $x_i$-direction, $g$ is the first-order derivative (with respect to any spatial coordinate) of any MHD variable, and $q_{max}$ is the maximum quenching factor. For any perturbation of the wavenumber $k$, this term quickly approaches $q_{max}$ as $k\rightarrow\infty$ and decreases with $k^2$ as $k\rightarrow 0$, hence ensuring that perturbations with a wavelength of same order as the grid size are heavily damped, while perturbations with wavelengths that are more than one order of magnitude larger than the grid size are only slightly damped. We used $q_{max}=8$ because this has been empirically shown to work well when Bifrost was used to solve standard test problems.

Thus, the hyper-diffusive resistivity of Bifrost can be written as a diagonal tensor, $\bar{\bar{\eta}}_{G11}$, given by
\begin{align}
    \eta_{G11,xx} &= \frac{\eta_3}{2} \left[
    U_{m,y} \Delta y \mathbb{Q}_y\left( \frac{\partial B_{z}}{\partial y} \right) 
    + U_{m,z} \Delta z \mathbb{Q}_z\left( \frac{\partial B_{y}}{\partial z} \right)
    \right] , \nonumber \\
    \eta_{G11,yy} &= \frac{\eta_3}{2} \left[
    U_{m,z} \Delta z \mathbb{Q}_z\left( \frac{\partial B_{x}}{\partial z} \right) 
    + U_{m,x} \Delta x \mathbb{Q}_x\left( \frac{\partial B_{z}}{\partial x} \right)
    \right] , \nonumber \\    
    \eta_{G11,zz} &= \frac{\eta_3}{2} \left[
    U_{m,x} \Delta x \mathbb{Q}_x\left( \frac{\partial B_{y}}{\partial x} \right) 
    + U_{m,y} \Delta y \mathbb{Q}_y\left( \frac{\partial B_{x}}{\partial y} \right)
    \right] , \nonumber \\ 
    \eta_{G11,xy} &= \eta_{G11,yx} = \eta_{G11,yz} = \eta_{G11,zy}= \eta_{G11,xz} = \eta_{G11,zx} = 0 . \label{eq:eta-hdiff}
\end{align}

This resistivity model ensures that the resistive terms in the induction and energy equation become significant only in the regions in which the diffusive velocity is high because of the high fast-mode velocity, advective velocity, or strong magnetic shocks along with strong gradients in the magnetic field, which allow the Reynolds number to stay high outside these regions.

\subsubsection{{\syntelis} model}\label{sec:s19-model}
The {\syntelis} resistivity, $\eta_{S19}$, is a scalar function given by

\begin{align}
    \eta_{S19} &= \left\{ 
    \begin{array}{ll}
        \eta_0, &  |{\bf J}| < J_{crit} \\
        \eta_0 + \eta_1 |{\bf J}|/J_{crit}, &  |{\bf J}| \geq J_{crit} \\
    \end{array}
    \right.  . \label{eq:anomalous_resistivity_syntelis} 
\end{align}
\noindent
\citet{2019ApJ...872...32S} used $\eta_0 = 3.78\times10^{-2}\ \mathrm{km^2\ s^{-1}}$, $\eta_1 = 3.78\times10^{-1}\ \mathrm{km^2\ s^{-1}}$, and $J_{crit}=5.00\times10^{-4}\ \mathrm{G\ km^{-1}}$.
In our experiments, we instead chose  $\eta_1 = 7.56\ \mathrm{km^2\ s^{-1}}$
in order to obtain approximately the same inflow Alfvén Mach number as when applying the {\hdiff} model on the 2D flux cancellation experiment (Sect. \ref{sec:Results-2Dflux}), as well as an average current sheet length similar to that of \citet{2019ApJ...872...32S}. This change was needed because the MHD solver scheme of Bifrost and the Lare3D code employed by \citet{2019ApJ...872...32S} are different. The Lare3D code is a Lagrangian-Eulerian Remap code \citep{2001JCoPh.171..151A}.

\subsubsection{{\yokshi} model} \label{sec:ys94-model}

The {\yokshi} resistivity, $\eta_{_{YS94}}$, is defined as
\begin{align}
    \eta_{_{YS94}} &= \left\{ 
    \begin{array}{ll}
        0, &  v_d \leq v_c \\
        \min(\alpha (\frac{v_d}{v_c} - 1)^2,\eta_{max}), &  v_d > v_c\\
    \end{array}
    \right. , \label{eq:anomalous_resistivity_yokoyama}
\end{align}
where $v_d = \frac{J}{n_e e}$ is the electron drift velocity, and $v_c$, $\alpha$, and $\eta_{max}$ are free parameters. \citet{1994ApJ...436L.197Y} used $v_c \in [4.16\times 10^{-7}, 8.32\times 10^{-6}]\ \mathrm{km\ s^{-1}}$, $\alpha \in [0.20, 2000]\ \mathrm{km^2\ s^{-1}}$, and $\eta_{max}=2000\ \mathrm{km^2\ s^{-1}}$ (normalisation units and formulae are extracted from \citealp{1992ApJS...78..267N} and \citealp{1996PASJ...48..353Y}).

In our simulations, we used $v_{c} = 8.3\times 10^{-6}\ \mathrm{km\ s^{-1}}$, $\alpha = 4.0\times 10^{-8}\ \mathrm{km^2\ s^{-1}}$, and $\eta_{max}=2000\ \mathrm{km^2\ s^{-1}}$ in order to obtain a similar inflow Alfvén Mach number in the 2D flux cancellation simulation as when using the other resistivity models.  With this, we applied a much lower value of the scaling factor $\alpha$ than \citet{1994ApJ...436L.197Y} used in their study of current sheets located in the convection zone. Our case deals with 
reconnection in current sheets that are located in the corona, where the density is several orders of magnitude lower. This causes the drift velocity in current sheets to become several orders of magnitude higher. It is therefore logical that a weaker scaling factor between resistivity and drift velocity is needed here. In addition to the resistivity given by Eq. \eqref{eq:anomalous_resistivity_yokoyama}, we added a background uniform resistivity of $\eta_0 = 4.00\times 10^{-2}\ \mathrm{km^2\ s^{-1}}$ when using this model, similar to that of {\syntelis}.


\subsection{Numerical experiments}
\label{sec:experiments}


\begin{figure}
    \centering
    \includegraphics[width=\columnwidth]{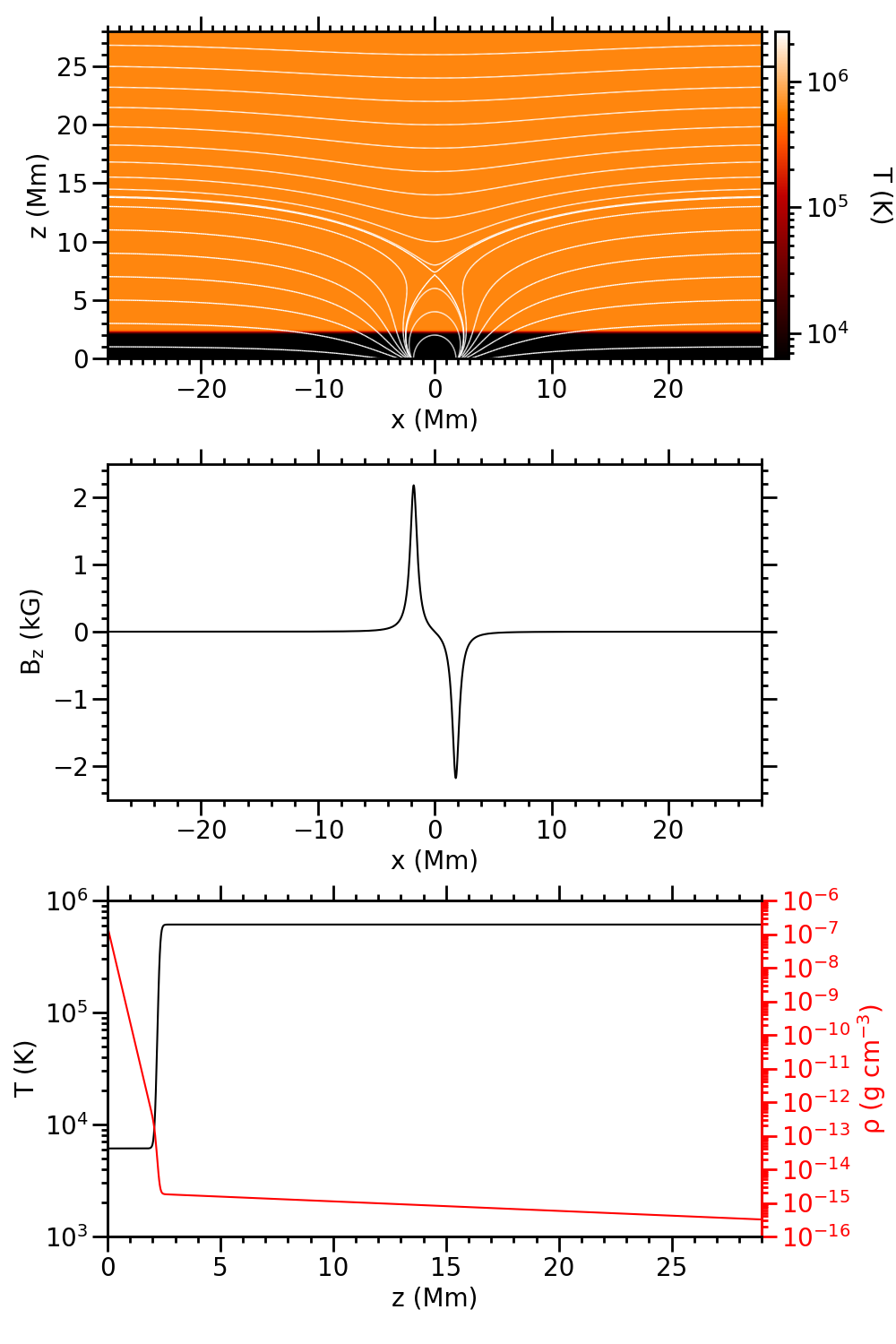}
    \caption{Initial conditions for the 2D flux cancellation experiment mimicking \cite{2019ApJ...872...32S}. Top: 
    Map of the temperature with the magnetic field topology superimposed.
    Middle: Vertical component of the magnetic field, $B_z$, at $z=0$. Bottom: Stratification of the temperature (black) and mass density (red).}
    \label{fig:init_syntelis}. 
\end{figure}

\subsubsection{2D flux cancellation experiment}
\label{sec:syntelis}
The first experiment mimics the case 1 simulation by \citet{2019ApJ...872...32S}, in which reconnection is driven by converging opposite polarities at the solar surface, leading to flux cancellation in a 2D atmosphere. The computational domain was given by $x \in [-30,30]$ Mm and $z \in [0,30]$ Mm, and it was discretised over $2048 \times 1024$ grid points.
The initial magnetic field was a superposition of two sources of opposite polarity placed below the photosphere, along with a horizontal uniform background magnetic field. In 2D, the magnetic field strength from one source with a flux of $F$ at a given distance ${\bf r} $ is $F/(\pi r),$ with the direction given by unit vector $\hat{\bf r} = {\bf r}/r$. Thus, the initial magnetic field is given by

\begin{align}
    {\bf B}(x,z,t=0) = \frac{F}{\pi} \frac{{\bf r}_1}{r_1^2} - \frac{F}{\pi} \frac{{\bf r}_2}{r_2^2} - B_0 \hat{\bf x} ,
    \label{eq:init_B_syntelis}
\end{align}
where $F = 2500\ \mathrm{G\ Mm}$ is the flux of each source, $B_0 = 45\ \mathrm{G}$ is the magnetic field strength of the horizontal background field, and
\begin{align}
    {\bf r}_1 &= (x+d_s) \hat{\bf x} + (z-z_0) \hat{\bf z} , \\
    {\bf r}_2 &= (x-d_s) \hat{\bf x} + (z-z_0) \hat{\bf z} ,
\end{align}
where $d_s = 1.8$ Mm is the initial half-separation distance between the sources, and $z_0=-0.36$ Mm is the height at which the sources are located.

The initial temperature profile of \citet{2019ApJ...872...32S}, set to mimic the C7 model of \citet{2008ApJS..175..229A}, is given by
\begin{align}
    T(x,z,t=0) = T_{pho} + \frac{T_{cor} - T_{pho}}{2}\left[ \tanh \left(\frac{z-z_{cor}}{w_{tr}}\right)  + 1 \right]  .\label{eq:init_T_syntelis}
\end{align}
with $T_{pho} = 6109\ \mathrm{K}$ and $T_{cor} = 0.61\ \mathrm{MK}$. For the location of the bottom of the corona and the width of the transition region, we used $z_{cor} = 2.31\ \mathrm{Mm}$ and $w_{tr} = 0.09\ \mathrm{Mm}$ in our simulations. The initial mass density was found by requiring hydrostatic equilibrium, $\partial P/\partial z =  -\rho g$, and a photospheric density of $\rho_{pho} = 1.67\times 10^{-7}\ \mathrm{g\ cm^{-3}}$. With $P$ given by the ideal gas law and $T$ given by Eq.~\eqref{eq:init_T_syntelis}, the following analytical solution was found:
\begin{align}
    \rho(x, \tilde z, t=0) = \rho_{pho} e^{-2 \chi_0 (\tilde{z}+\tilde{z}_c)} \left( \frac{ T_{pho} + T_{cor} e^{2 \tilde{z}} }{ T_{pho} + T_{cor} e^{-2 \tilde{z}_c} } \right)^{\chi_0 -  \chi_1}  \frac{T_{pho}}{T( \tilde z)} ,\label{eq:init_rho_syntelis}
\end{align}
where
\begin{align}
    \tilde{z} \equiv \frac{z-z_{cor}}{w_{tr}}, \quad \tilde{z}_c \equiv \frac{z_{cor}}{w_{tr}}, \quad \chi_0 \equiv \frac{\mu m_H gw_{tr}}{2k_B T_{pho}}, \quad \chi_1 \equiv \frac{\mu m_H gw_{tr}}{2k_B T_{cor}} .
\end{align}

\noindent
Initial magnetic field, temperature, and mass density computed from the above equations are shown in Fig.~\ref{fig:init_syntelis}. The figure shows that the initial conditions of \citet{2019ApJ...872...32S} have indeed been successfully mimicked.

For the bottom boundary conditions, we used a driving mechanism where the horizontal velocity $u_x$ is defined as 
\begin{align}
    u_x(x,z=0,t) = \left\{ 
    \begin{array}{cc}
        v_0(t) & x<0 \\
        0 & x = 0 \\
        -v_0(t) & x>0
    \end{array}
    \right. , \label{eq:bound_ux_syntelis}
\end{align}
where
\begin{align}
    v_0 (t) = \frac{1}{2} v_{max} \left[ \tanh \left(\frac{t-t_0}{w}\right) + 1 \right] , \label{eq:v0_driver_syntelis} 
\end{align}
$v_{max} = 1\ \mathrm{km\ s^{-1}}$, $t_0 = 10.1$ minutes, and $w = 1.4$ minutes; and the
magnetic field ${\bf B}$ is given by
\begin{align}
    {\bf B}(x,z=0,t) = \frac{F}{\pi} \frac{{\bf r_1}(t)}{r_1^2(t)} - \frac{F}{\pi} \frac{{\bf r_2}(t)}{r_2^2(t)} - B_0 \hat{\bf x} ,\label{eq:bound_B_syntelis}
\end{align}
where 
\begin{align}
    {\bf r_1}(t) = (x+d(t)) \hat{\bf x} + (z-z_0) \hat{\bf z} , \label{eq:bound-r1}\\
    {\bf r_2}(t) = (x-d(t)) \hat{\bf x} + (z-z_0) \hat{\bf z} \label{eq:bound-r2},
\end{align}
and 
\begin{align}
    d(t) = d_s - \left( v_{max}\frac{w}{2} \left[ \ln \left( \cosh \left(\frac{t-t_0}{w}\right) \right) -\ln \left( \cosh \left(\frac{t_0}{w}\right) \right) \right] +\frac{1}{2}v_{max}t \right)
    \label{eq:bound-d_t}.
\end{align}
In addition, an absorbing layer was applied on $u_z$, $\rho$, and $e$ to ensure that waves hitting the boundaries were not reflected. With respect to the top boundary, we set $u_x=0$, ${\bf B}$ to be line-tied to the flow, and applied an absorbing layer for $u_z$, $\rho,$ and $e$.

Because Bifrost is designed to use periodic side-boundaries, 
we superimposed additional terms to the initial and bottom boundary conditions for ${\bf B}$, Eqs.~\eqref{eq:init_B_syntelis} and \eqref{eq:bound_B_syntelis}, which corresponds to magnetic sources located in neighbouring domains identical to our computational domain. This adjustment had a negligible effect on the central parts of the domain, where the reconnection takes place, but it ensured that the field was horizontal and $\nabla \cdot {\bf B}$-free at the periodic side-boundaries. For $u_x$, $\rho$, and $e$, we also applied an absorbing layer, thus keeping a periodic side-boundary.

As an additional note regarding the boundaries, the {\syntelis} and {\yokshi} resistivity models in this experiment were applied within $x \in [-28,28]$ Mm $\land\ z \in [2,28]$ Mm. 
The resistivity was set uniformly to $\eta_0$ outside these regions to avoid conflicts near the boundary layers.

\subsubsection{1D Harris current sheet}
\label{sec:harris}

Our second experiment was a 1D Harris current sheet that was set up in a computational domain of $z \in [-2, 2]$ Mm and was discretised over 4096 grid points. To keep this experiment relatively simple, we neglected the gravitational term, $\rho {\bf g}$, and the Spitzer thermal conductivity term, $Q_C$, when solving Eqs.~\eqref{eq:continuity}-\eqref{eq:energy}. The initial condition for the magnetic field was
\begin{align}
    {\bf B}(z, t=0) = B_0 \tanh \left( (z-z_0)/w  \right) \hat{\bf x} . \label{eq:init_B_harris}
\end{align}
When we assume a uniform total pressure (the sum of gas pressure and magnetic pressure) with a uniform temperature $T(z,t=0)=T_0$, the initial density is given by
\begin{align}
    \rho (z, t=0) = \rho_0 + \frac{\mu m_H}{k_B T_0}\frac{B_0^2}{\sqrt{8\pi}} \left(1 - \tanh^2 \left( (z-z_0)/w  \right) \right)  , \label{eq:init_rho_harris}
\end{align}
where $\rho_0$ is the density far away from the current sheet. In our simulations, we used $T_0=0.61$ MK, $\rho_0 = 10^{-15}\ \mathrm{g\ cm^{-3}}$, and $B_0=1$ G (as well as $w=20$ km and $z_0=0$) in order to approximately match the temperature, mass density, and magnetic field strength in the inflow region of the current sheet of the 2D flux cancellation experiment (Sect.~\ref{sec:syntelis}). This ensured that the Alfvén velocity and current density in the 1D and 2D experiment were of the same order of magnitude in the regions near the current sheets, which facilitated performing the same comparisons between the same resistivity models in the two experiments. 

The boundary condition was handled by applying an absorbing layer for all variables near the two boundaries to ensure that no waves hitting the boundaries were reflected back into the physical domain. The {\syntelis} and {\yokshi} resistivities on this experiment were applied within $z \in [-0.5,0.5]$ Mm and were set uniformly to $\eta_0$ elsewhere  to avoid conflicts near the boundary layers.



\section{Results}
\label{sec:Results}

\subsection{2D flux cancellation experiment}
\label{sec:Results-2Dflux}

\subsubsection{Overview}
\label{sec:Results-2Dflux-Description}

\begin{figure}
    \centering
    \includegraphics[width=\columnwidth]{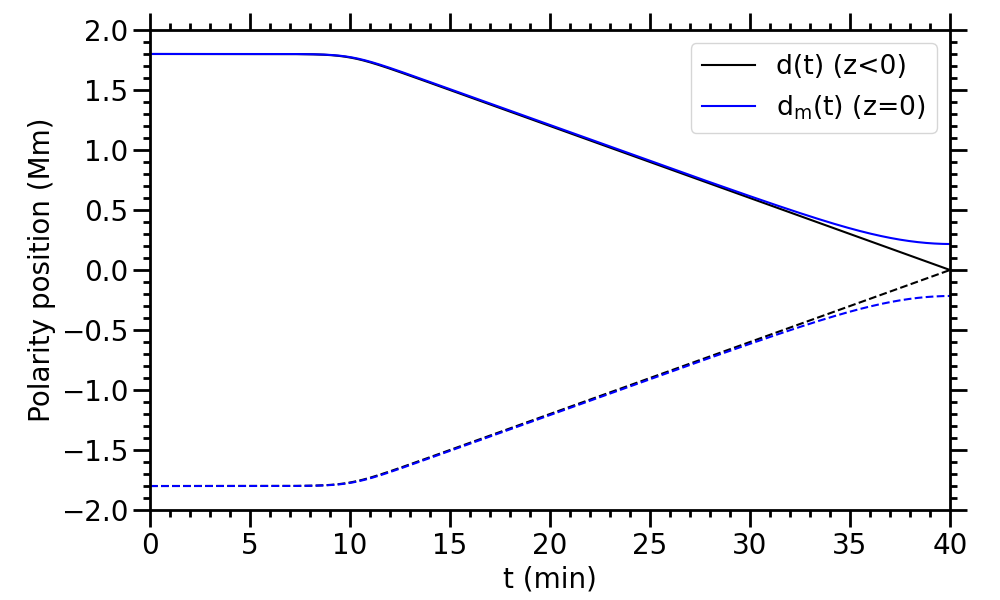}
    \caption{Evolution of the magnetic polarities in the 2D flux cancellation experiment. The black lines show the horizontal position of each source, given by Eq.~\eqref{eq:bound-d_t}. The blue lines show the horizontal position of the photospheric polarities, that is, the location along $z=0$ where $B_z$ reaches its maximum value. The results here are from the simulation with the {\hdiff} resistivity model, but nearly identical results are obtained with the other two resistivity models.}
    \label{fig:syntelis-fig5b}. 
\end{figure}

\begin{figure*}
    \centering
    \includegraphics[width=\textwidth]{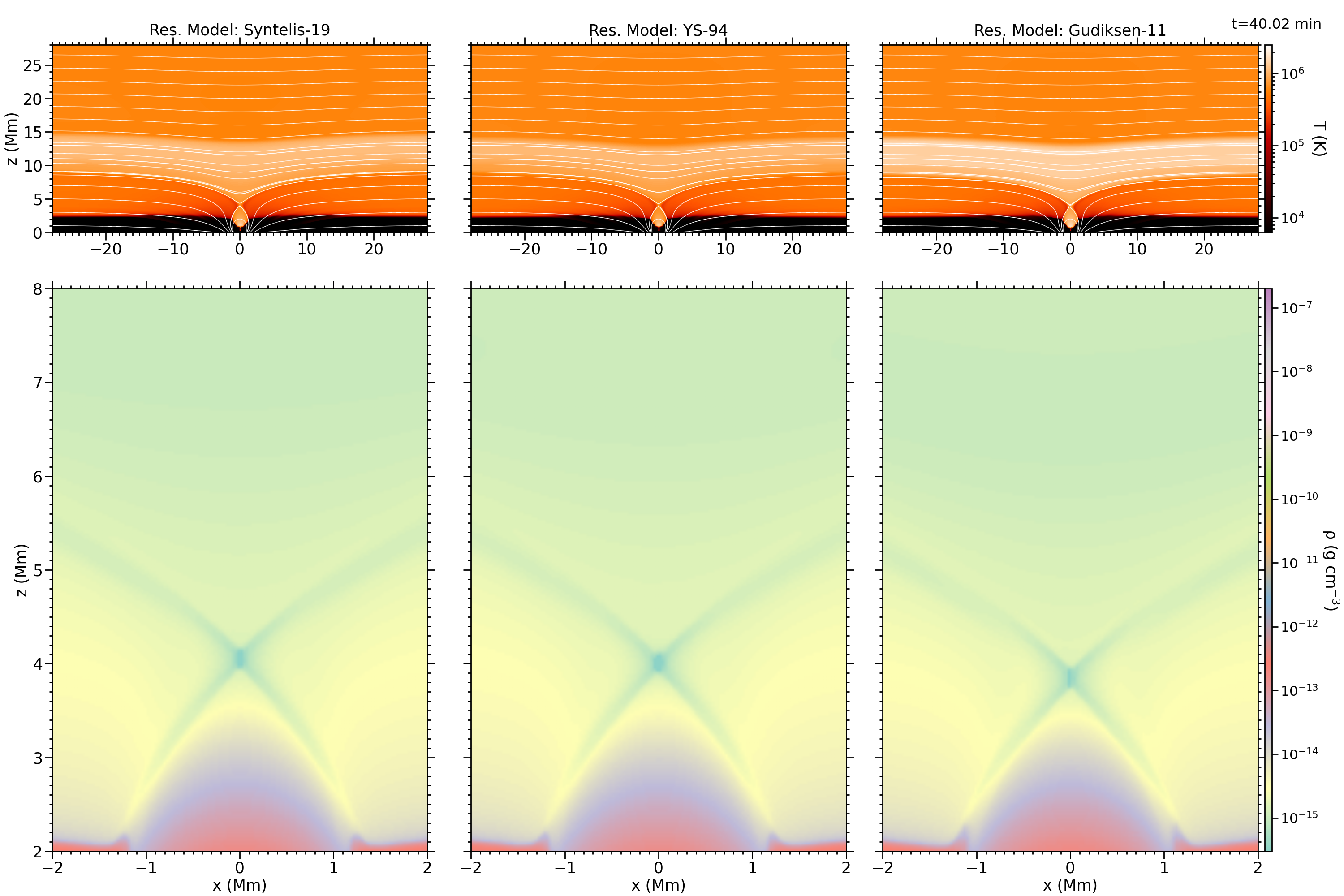}
    \caption{Atmospheric response in the 2D flux cancellation experiment for the three resistivity models (columns) at $t=40$ min. Top: Maps of the temperature with the magnetic field topology superimposed. Bottom: Maps of the mass density around the reconnection site. A movie of the full evolution from $t=0$ to $t=40$~min of these maps is available online.}
    \label{fig:syntelis-t=40}
\end{figure*}

The simulation was run for 40 min, and the results show that the large-scale evolution of the main quantities such as the magnetic field, temperature, and density agrees relatively well with the case 1 simulation of \citet{2019ApJ...872...32S} for the three resistivity models. 

The two sources of opposite magnetic polarity located immediately below the photosphere move towards each other with the driving velocity given by Eq. \eqref{eq:v0_driver_syntelis} until they meet at $x=0$ at $t=40$ min. Fig.~\ref{fig:syntelis-fig5b} shows that the above-lying photospheric polarities do indeed follow the driver very well throughout the simulation time, until they start to slow down after $t=35$ min, similar to \citet{2019ApJ...872...32S}.

As a consequence of the motion of the photospheric polarities towards each other, the null-point, initially located 7.6 Mm above the photosphere, is stretched into a vertical current sheet with a length of up to $\sim 0.6$ Mm. The reconnection site moves slowly downwards along $x=0$ during the cancellation phase, that is, from $t=10$ min to $t=40$ min. Thermal energy from the reconnection is transported outwards from the current sheet along the magnetic field lines and heats up a wide nearly horizontal open reconnection loop above it and a narrow closed reconnection loop below it.
The top panels of figure \ref{fig:syntelis-t=40} show maps of the temperature in the atmosphere at $t=40$ min for each resistivity model. The magnetic field topology is superimposed. The bottom panels show the corresponding maps of the mass density in the region surrounding the null-point.\footnote{These plots mimic the style of \citet{2019ApJ...872...32S} to facilitate comparison.} 
The resistivity models are indeed capable of producing a large-scale atmospheric response that agrees among the models, except for some differences in terms of final null-point height and maximum temperature. The height of the elongated null-point (here defined as the centre of the current sheet) at $t=40$ min lies at 4.05 Mm above the photosphere in the {\syntelis} case, 4.0 Mm in the {\yokshi} case, and 3.85 Mm in the {\hdiff} case. The maximum temperature in the heated region at this time is 1.49 MK in the {\syntelis} case, 1.38 MK in the {\yokshi} case, and 1.78 MK in the {\hdiff} case.

A movie of Fig.~\ref{fig:syntelis-t=40} is available online. It shows the evolution of the temperature, magnetic field, and density throughout the whole simulation time for the three cases. While all cases eventually have temperature profiles of the same structural shape, despite some differences in terms of maximum temperature and null-point height, the plasma inside the current sheet behaves notably differently in each case. In the {\syntelis} model, the current sheet moves steadily downwards without any sign of plasmoid generation. In the other two resistivity models, plasmoids are generated rapidly. The current sheet in the {\yokshi} case is different from the other two cases by its remarkably lower mass density. In the {\hdiff} case, the current sheet coincides with a thin stripe of increased mass density. This is also visible in the {\syntelis} case, but to a lesser extent. 

The Lundquist number at the centre of the current sheet is
$\sim 5$ in the {\hdiff} case, $\sim 10$ in the {\syntelis} case, and $\sim 20-100$ in the {\yokshi} case, while the Reynolds number inside the current sheet approaches unity in all three cases (but it is slightly higher in the {\yokshi} case). At a horizontal distance of 0.1 Mm from the current sheet, the Reynolds and Lundquist numbers are $\sim 10^4$ or higher in all three models. This is as expected because the resistivity models were scaled so that the simulation was able to obtain roughly the same Alfvén velocities in the inflow region.
The plasma-$\beta$ inside the current sheet reaches maximum values (in the top and bottom points of the current sheet) of $\sim 2-5$ in the {\hdiff} case, $\sim 1$ in the {\syntelis} case, and $\sim 0.5$ in the {\yokshi} case. At a distance of 0.1 Mm from the current sheet, $\beta \sim 0.1$ in all three cases.

\begin{figure*}
    \centering
    \includegraphics[width=\textwidth]{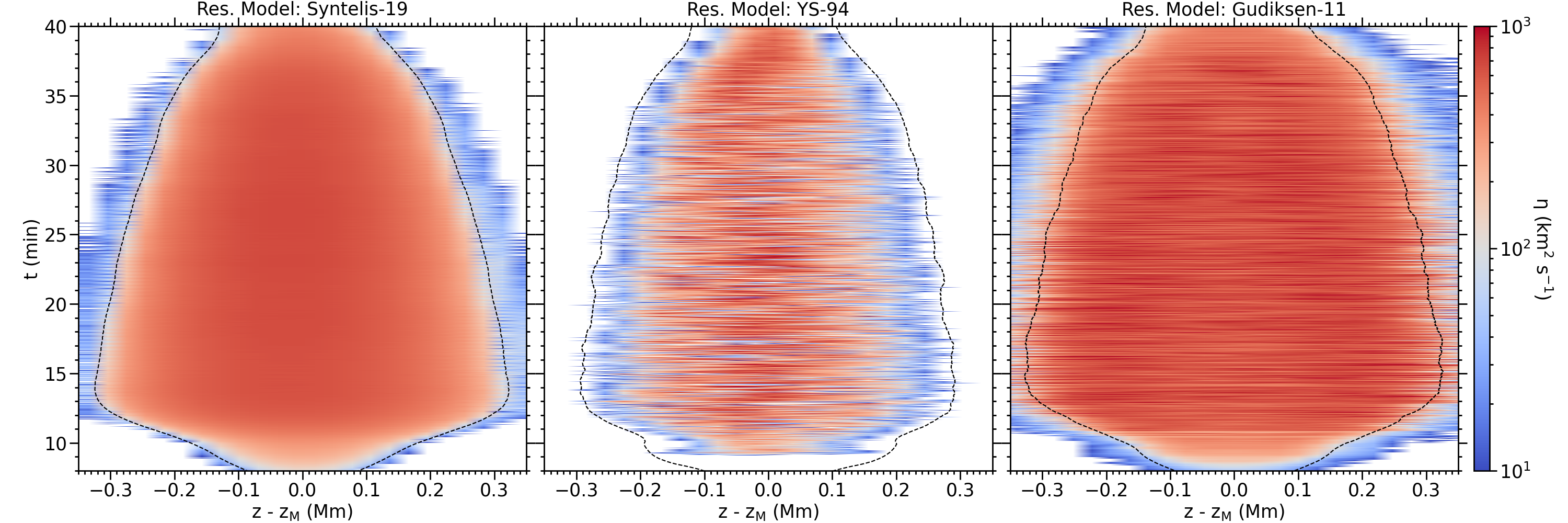}
    \caption{Evolution of the resistivity $\eta$ along the current sheet for each resistivity model. The resistivity is measured at $x=0$ and is shown as function of the height relative to the current sheet midpoint, $z_M$. The dashed lines mark the top and bottom of the current sheet, which are annotated as $S_h$ and $S_l$, respectively, in Fig.~\ref{fig:cs-snapshot-tg}.}
    \label{fig:results-2dflux-etaplots}
\end{figure*}

To demonstrate that the three resistivity models work differently on the current sheet, maps of the resistivity along $x=0$ as function of height relative to the vertical midpoint of the current sheet and time for each resistivity model are shown in Fig.~\ref{fig:results-2dflux-etaplots}. The dashed lines in each panel mark the top and bottom of the current sheet. The relatively smooth behaviour of the resistivity of the {\syntelis} model agrees well with the fact that the current sheet in this case evolves steadily without any sign of plasmoid instability. Based on this, it is plausible to expect the current sheet in this case to follow a Petschek-like reconnection scheme, especially in terms of energy conversion, which is analysed in Sect.~\ref{sec:results-2dflux-energy}. The resistivity of the {\hdiff} and {\yokshi} models, on the other hand, varies more rapidly in its magnitude due to the frequent plasmoid generation, and therefore we expect the energy conversion rates in these cases to deviate more significantly from the Petschek theory. While the {\hdiff} and {\syntelis} resistivities inside the current sheet mostly stay within the range of $100$ to $1000 \ \mathrm{km^2\ s^{-1}}$, the  {\yokshi} resistivity has a lower average value that reaches below $100\ \mathrm{km^2\ s^{-1}}$ within the boundaries of the current sheet. Along with the fact that the diffusive layer is shorter than in the other cases, this explains why the atmosphere in this case has the lowest maximum temperature: the Joule heating scales directly with the resistivity. Although the diffusive layer in the {\hdiff} case is of similar size as in the {\syntelis} case, the average resistivity of the current sheet in the {\hdiff} case is slightly higher because the resistivity is enhanced in the plasmoids that appear relatively frequently. This explains why the atmosphere receives the highest amount of heating in the {\hdiff} case.

\subsubsection{Comparison method}
\label{sec:Results-2Dflux-Comparison}

\begin{figure*}
    \centering
    \includegraphics[width=\textwidth]{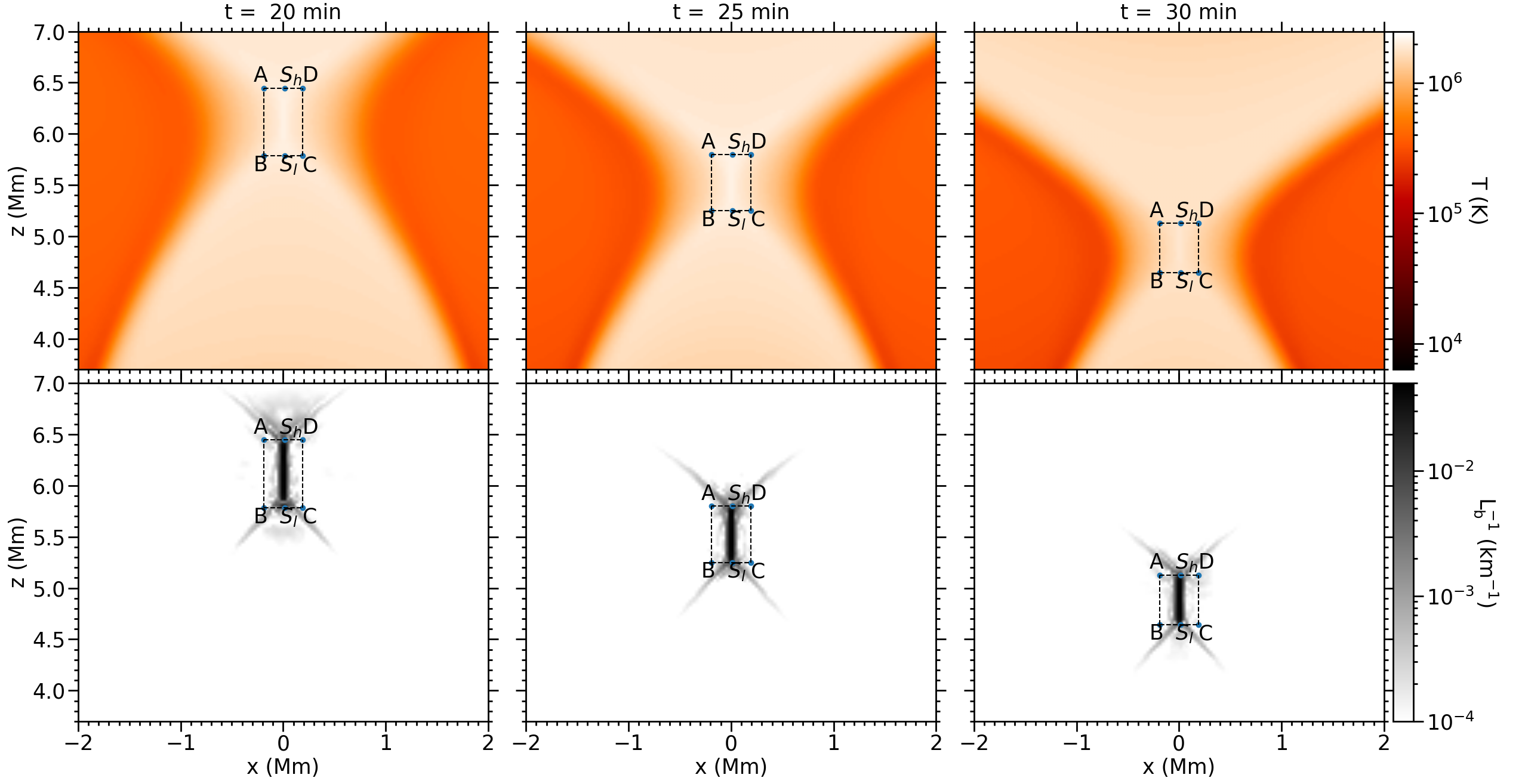}
    \caption{Evolution of the current sheet in the 2D flux cancellation experiment for the {\hdiff} resistivity model at different times (columns). Top: Temperature maps zoomed in on the region around the elongated current sheet. Bottom: Corresponding maps of the inverse characteristic length $L_B^{-1}$. The rectangle marks the region around the current sheet, and  the inflow parameters are measured at the line segments AB and CD. $S_h$ and $S_l$ are the top and bottom of the current sheet, respectively.}
    \label{fig:cs-snapshot-tg}
\end{figure*}

We performed the same comparison between simulations and theory as \citet{2019ApJ...872...32S} by locating the current sheet 
and measuring some inflow values near it and comparing them with values predicted from analytical formulae.
To demonstrate the localisation of the current sheet and the regions in which the inflow values are measured, Fig.~\ref{fig:cs-snapshot-tg} shows maps of the temperature and inverse characteristic length of the magnetic field, $L_B$, in the surroundings of the null-point with the inflow region delimited by a rectangle of points A, B, C, and D. We defined the current sheet as the oblong vertical region along $x=0$ where the characteristic length for magnetic field, $L_b$, is shorter than a chosen threshold value of 100 km, which is roughly three grid cells because the numerical resolution of the experiments is $\sim 30$ km. The extremes of the current sheet are indicated in the plots with $S_h$ (top) and $S_l$ (bottom). The corresponding current sheet length $L_m$ is measured as the vertical distance between these two points. The index $m$ denotes that it is a numerically measured value. This indexation was applied to several numerically measured values in order to distinguish them from their analytical counterparts. Points A, B, C, and D are defined such that the AB and CD segments form vertical lines parallel to the current sheet at 0.2 Mm to the left and right of the current sheet, respectively.
The choice of this location of the line segments was made so that the segments lay within the range  in which the analytical formulae for the inflow values used by \citet{2019ApJ...872...32S} are valid. We found that placing AB and CD at any horizontal distance between 0.1 and 0.2 Mm was suitable. We used 0.2 Mm to also be consistent with the criterion employed by \citet{2019ApJ...872...32S}.
The figure shows that the inflow rectangle ABCDA does indeed follow the current sheet as it moves downwards throughout the cancellation phase.
 
The inflow magnetic field strength $B_{im}$ and velocity $v_{im}$ were measured as the mean absolute value of the magnetic field and the velocity, respectively, along the line segments AB and CD. The Poynting influx $\Phi_{S_{im}}$ was measured by integrating the Poynting vector component perpendicular to these line segments, $S_x = [{\bf E}\times {\bf B}]_x/\mu_0 = E_y B_z / \mu_0$, over AB and CD. The average density along AB and CD, $\rho_{im}$, was also measured because it is needed in the calculations of the analytical estimate for the Poynting influx.

Knowing the numerical measures $B_{im}$, $v_{im}$, $L_m$, and $\Phi_{S_{im}}$, 
we compared them with analytical estimates for $B_{i}$, $v_{i}$, $L$, and $\Phi_{S_{i}}$, as derived by \citet{2019ApJ...872...32S}.
The analytical expression for the inflow magnetic field strength $B_{i}$ is
\begin{align}
    B_i (d, d_0, L) &= B_0 \sqrt{\frac{d_0}{d} - 1} \frac{L}{d_0} , \label{eq:inflow-B}
\end{align}
where $d$ and $d_0$ are the source separation distance and the critical source separation distance, respectively. Two different analytical estimates were made for $B_i$: 
1) $B_i (d(t), d_0, L_m)$, based on the source positions with $d(t)$  given by Eq.~\eqref{eq:bound-d_t} and $d_0 = \frac{2F}{\pi B_0}$ ;
and 2) $B_i (d_m(t), d_{0m}, L_m)$, based on the photospheric polarity positions, where $d_m(t)$ is the half-separation distance between the photospheric polarities, shown as the blue curve in Fig.~\ref{fig:syntelis-fig5b}, and $d_{0m} = \frac{2F_m}{\pi B_0}$, where $F_m = 2\ 200$ G Mm is the flux of each photospheric polarity.

The analytical expression for the inflow velocity is
\begin{align}
    v_i (v_0, d_0, L) &= f (d, d_0, z_0) \ v_0\frac{d_0}{L} , \label{eq:inflow-v} 
\end{align}
where
\begin{align}
    f (d, d_0, z_0) = 1 - d_0 \frac{z_{max} - z_0}{(z_{max} - z_0)^2 + d^2} \frac{1}{\sqrt{d_0/d - 1}}  
\end{align}
is a flux correction factor, as explained in detail in the appendix of \citet{2019ApJ...872...32S}, with $z_{max} = 30$ Mm as the top of the computational domain. The factor was initially $f\approx 0.72$ when $d=1.8$ Mm, then approached 1 as $d\rightarrow 0$. Again, two analytical estimates were made 
for the inflow velocity: 
1) $v_i (v_0(t), d_0, L_m)$, based on the sources, with $v_0(t)$ given by Eq.~\eqref{eq:v0_driver_syntelis}; 
and 2) $v_i (v_{0m}(t), d_{0m}, L_m)$, based on the photospheric polarities, using $f(d_m, d_{0m}, 0)$, and where $v_{0m}(t)\equiv \dot{d}_m(t)$ is the absolute value of the velocity of the photospheric polarities given by the time derivative of the blue curve in Fig.~\ref{fig:syntelis-fig5b}.

The analytical current sheet length is
\begin{align}
    L (M_A, d, d_0, v_0, v_{A0}) &= \sqrt{f(d, d_0, z_0)} \ d_0 \sqrt{\frac{M_{A0}}{M_A} \frac{1}{\sqrt{d_0/d - 1}}}  , \label{eq:cslength}
\end{align}
where $M_A$ is the inflow Alfvén Mach number, and $M_{A0} \equiv v_0/v_{A0}$ is a hybrid Alfvén Mach number based on the hybrid Alfvén speed $v_{A0}\equiv B_0/\sqrt{\mu_0 \rho_i}$, a quantity introduced by \citet{2019ApJ...872...32S} which is based on the external magnetic field $B_0$ but the inflow mass density $\rho_i$ (therefore "hybrid"). We estimated 1) $L (M_{Am}, d(t), d_0, v_0(t), v_{A0m})$ based on sources, with $v_{A0m} = B_0/\sqrt{\mu_0 \rho_{im}}$, and 2) $L (M_{Am}, d_m(t), d_{0m}, v_{0m}(t), v_{A0m})$ based on photospheric polarities.

\noindent
The analytical Poynting influx is
\begin{align}
\Phi_{S_i} (M_A, d, d_0, v_0, v_{A0}) &= 2 f^2(d, d_0, z_0) \ \frac{v_0 B_0^2}{\mu_0} d_0 \sqrt{d_0 / d - 1} \frac{M_{A0}}{M_A} , 
\label{eq:Poynting-influx}
\end{align}
where we estimated 1) $\Phi_{S_i} (M_{Am}, d(t), d_0, v_0(t), v_{A0m})$ based on the sources and 2) $\Phi_{S_i} (M_{Am}, d_m(t), d_{0m}, v_{0m}(t), v_{A0m})$ based on the photospheric polarities.

We also calculated the fractions of the Poynting influx that were converted into kinetic energy and into heat. According to Gauss' theorem, we have

\begin{align}
    \Phi_{S_i} \equiv \oint_C \frac{1}{\mu_0}  {\bf E}\times{\bf B} \cdot d{\bf C} = \int_A \frac{1}{\mu_0} \nabla \cdot ({\bf E}\times{\bf B}) dA ,
\end{align}
where $C$ is the curve over the points $ABCDA,$ and $A$ is its enclosed area. This simply states that the energy increase in the system equals the energy added into it. The above equation can, with the help of vector calculus as well as  Faraday's law, Ohm's law, and Ampère's law, be rewritten as
\begin{align}
    \left|\Phi_{S_i}\right|  \approx \left|\int_A \eta {\bf J}^2 dA \right|  + \left|\int_A {\bf J}\cdot ({\bf v}\times {\bf B}) dA \right| ,
\end{align}
which indeed tells us that the input magnetic energy is converted into heat (first right-hand-side term) and kinetic energy (second right-hand-side term) through reconnection.
A third right-hand-side term, $\int_A \frac{\partial}{\partial t}\left(\frac{B^2}{2 \mu_0} \right)$, was neglected here as \cite{2019ApJ...872...32S} did the same (we measured this term in our simulations, and it is indeed small compared to the other right-hand side terms in the above equation). To compare the simulated energy conversion with \citet{1964NASSP..50..425P} theory, we measured the ${\bf J}\cdot ({\bf v}\times {\bf B})$ term and the Joule heating term integrated over the rectangle $A$ and compared it to three-fifths and two-fifths of the Poynting influx, respectively. For this comparison, we used both the numerical measure $\Phi_{S_{im}}$ and the analytical estimate $\Phi_{S_i}$.

\begin{figure*}
    \centering
    \includegraphics[width=\textwidth]{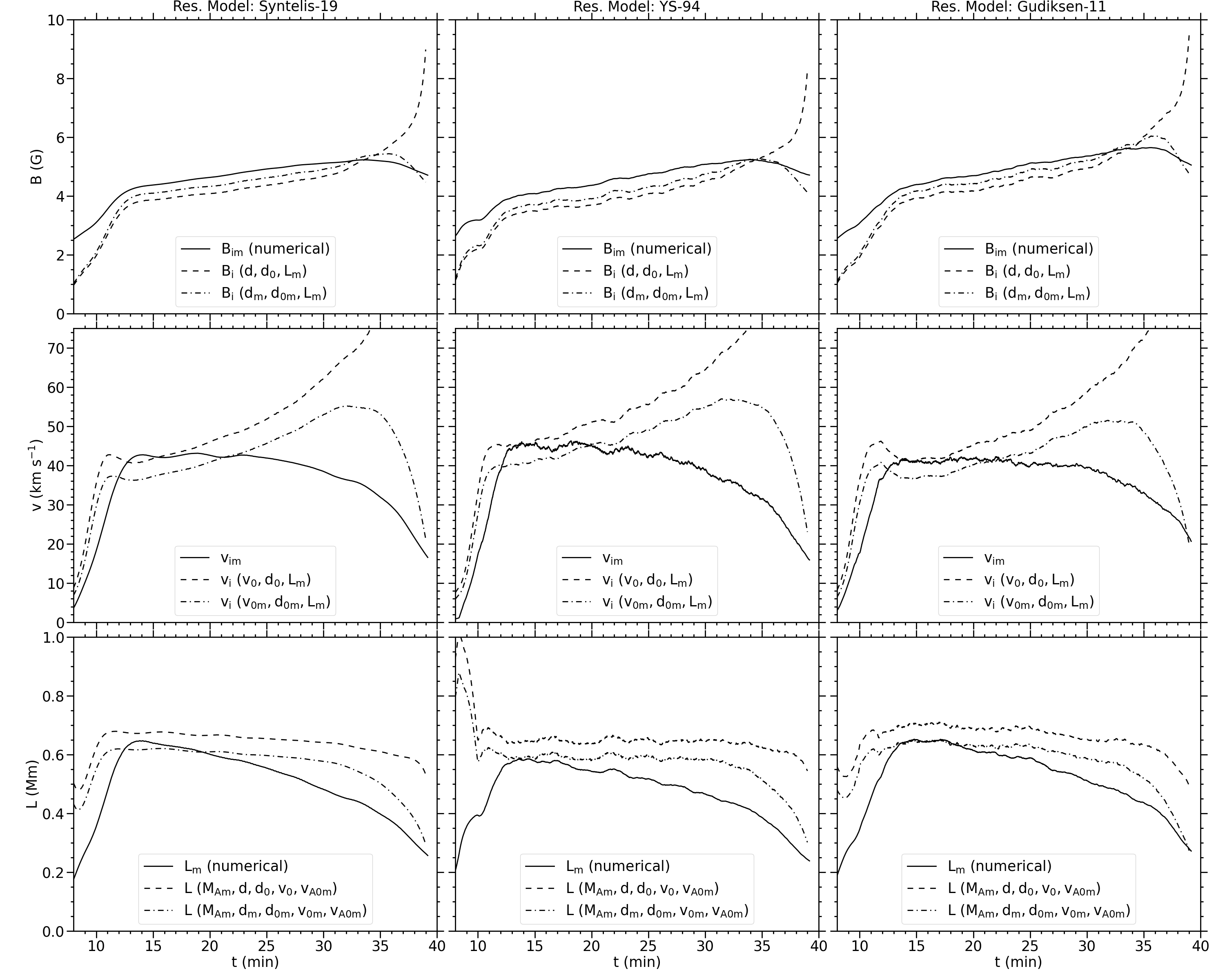}
    \caption{Evolution of the relevant quantities in the 2D flux cancellation experiment for each resistivity model (columns). Top: Inflow magnetic field, both numerical measures (solid curves) and analytical estimates (dashed and dash-dotted curves). Middle: Inflow velocity. Bottom: Length of the current sheet. The quantities are averaged over 100 s to reduce their rapid fluctuations.}
    \label{fig:cs-analysis1}
\end{figure*}

\subsubsection{Inflow magnetic field, velocity, and current sheet length}

Figure \ref{fig:cs-analysis1} shows the comparison between the numerical results (solid lines) and the analytical estimates based on the dynamics of the sources (dashed curves) and the photospheric polarities (dash-dotted curves) for the inflow magnetic field (top panels), the inflow velocity (middle panels), and the current sheet length (bottom panels). The quantities shown in the figure were averaged over 100 s to obtain smooth lines, which reduced their rapid fluctuations as a consequence of the non-stationary nature of the current sheet.

 It is clear from the figure that the numerical measures for $B_{im}$, $v_{im}$, and $L_m$ 
 in each model satisfactorily agree with each other and with the analytical estimates, especially those based on the photospheric response (dash-dotted curves), but they are not identical. The current sheet length in the {\yokshi} case is slightly shorter than in the other cases, which means that it deviates more strongly from the analytical estimate.  The current sheet length in the {\syntelis} case is similar to that of the {\hdiff} case in the first 10 minutes of the cancellation phase, but it then declines faster. The agreement is best in the {\hdiff} model for the numerical measure for $L_m$ and the analytical estimate for $L$ based on photospheric polarities. The inflow velocity in the {\syntelis} case is more or less the same as in the {\yokshi} case, both numerically and analytically, while the inflow velocity in the {\hdiff} case has a lower maximum value, and the numerical measure and the analytical estimate based on photospheric polarities agree better. The inflow magnetic field in the {\hdiff} case has a slightly higher maximum field strength than in the other two cases and simulation and theory agree best, while the field strength in the {\yokshi} case is weakest and simulation and theory deviate most.
 
The analytical estimates for  $L$ in each model agree very well with each other from $t=15$ min and throughout the simulations because the Alfvén Mach number, on which the analytical current sheet length is directly dependent, agrees well. We adjusted the input values of the diffusion scaling parameters of each model ($\eta_1$ for the {\syntelis} model, $\alpha$ for the {\yokshi} model, and $\eta_3$ for the {\hdiff} model) on purpose in order to obtain this agreement between the analytical estimates. The analytical estimates for $B_i$ and $v_i$ agree less well when comparing the resistivity models because these estimates depend on the numerical measures for $L_m$, which are slightly different in each case.

.

\begin{figure*}
    \centering
        \includegraphics[width=\textwidth]{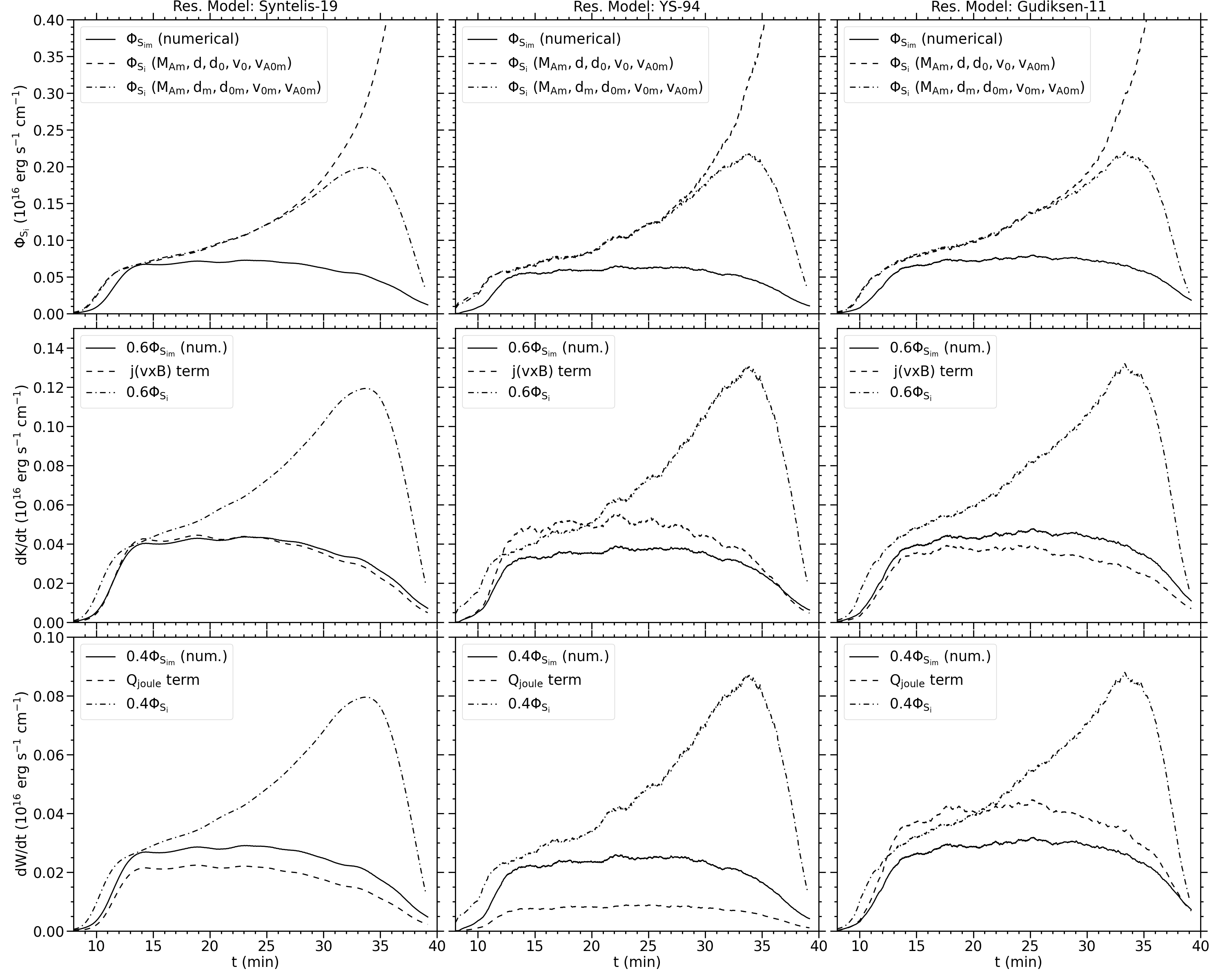}
    \caption{Evolution of the energy release in the 2D flux cancellation experiment for each resistivity model (columns). Top: Poynting influx, both numerical measures (solid curves) and analytical estimates (dashed and dash-dotted curves). Middle: Three-fifths of the numerical measure for the released energy (solid lines), compared to the numerical value for the kinetic energy output (dashed curves) and three-fifths of the analytical estimate for the released energy (dash-dotted curves). Bottom: Two-fifths of the numerical measure for the released energy (solid lines), compared to the numerical value for the heat output (dashed curves) and two-fifths of the analytical estimate for the released energy (dash-dotted curves). The quantities are averaged over 100 s to reduce their rapid fluctuations.}
    \label{fig:cs-analysis1-energy}
\end{figure*}

\subsubsection{Energy release} \label{sec:results-2dflux-energy}
Figure \ref{fig:cs-analysis1-energy} shows the energy release in the three models. The quantities here are also averaged over 100 s to reduce their rapid fluctuations. The first row shows the numerical measures of the Poynting influx $\Phi_{S_{im}}$ (solid line), and the analytical estimates for $\Phi_{S_{i}}$ based on the source positions (dashed curve) and based on the photospheric polarity positions (dash-dotted curve). 
In all three cases, the numerical measures approach the analytical estimate at $t=13$ min, which is approximately the time at which the current sheet length reaches its maximum value. After this time, the numerical Poynting influx stays constant in each case for the next 15 min, instead of increasing, as analytically predicted, before it slowly decreases. These numerical measures roughly follow the same evolution in all the three cases, however, but they reach a slightly lower maximum value in the {\yokshi} case, and are roughly of same order of magnitude as the analytical estimates based on photospheric polarities.

The second and third rows show the fraction of the energy that is released through reconnection that is transformed into kinetic energy and thermal energy, respectively, compared to three-fifths and two-fifths, respectively, of the numerical measures and analytical estimates for the Poynting influx. The energy conversion with the {\syntelis} model is more Petschek-like than with the other two models, with almost exactly three-fifths of the energy input converted into kinetic energy, and slightly less than two-fifths converted into heat. In the {\hdiff} model, significantly more than two-fifths of the input energy is converted into heat. It gains more heat than the other two models, and therefore, the agreement between the numerically measured and analytically predicted heat output is best. The {\yokshi} model deviates most from the Petschek theory: less than one-fifth of the energy is converted into heat.
This agrees with Fig.~\ref{fig:syntelis-t=40}, in which the {\hdiff} case resulted in the warmest atmosphere. The maximum temperature was almost 0.3 MK higher than in the {\syntelis} case, while the {\yokshi} model had the coldest atmosphere with a maximum temperature 0.1 MK lower than in the {\syntelis} case.

The {\syntelis} case follows a nearly perfect Petschek-like energy conversion. This agrees with the fact that this simulation has nearly no sign of plasmoid generation in the current sheet, as seen in the movie of Fig.~\ref{fig:syntelis-t=40}. This means that this resistivity model allows the current sheet to undergo Petschek reconnection. In the {\yokshi} and {\hdiff} models, the current sheet undergoes plasmoid-mediated reconnection, which explains why the kinetic and thermal energy released through reconnection is not necessarily equal to three-fifths and two-fifths, respectively, of the input magnetic energy. Still, it is noteworthy that these two cases, while they are plasmoid-mediated, follow completely different energy conversion schemes. While in the {\hdiff} case, more of the magnetic energy is converted into heat than predicted with Petschek theory and less into kinetic energy, in the {\yokshi} case, less magnetic energy is converted into heat and more into kinetic energy. 
As we described above, this is caused by the significantly stronger diffusive layer in the Gudiksen-11 model than in the {\yokshi} model, as shown in Fig.~\ref{fig:results-2dflux-etaplots}, where the {\hdiff} model clearly has the highest mean resistivity along the centre of the current sheet.
The frequency of plasmoids in current sheets as a result of different resistivity models and how this affects the heating of the surrounding plasma will be studied more in detail in an upcoming paper.

\subsubsection{Dependence on the choice of diffusion parameters} \label{sec:results-2dflux-freeparameters}
The results of the above section were obtained by setting the free parameters of the resistivity models to specific values to ensure that the inflow Alfvén speed has roughly the same value in all simulation cases. In this way, we ensured that we solved a very similar physical problem even though we used different numerical approaches. In this section, we study the dependence of the results on an adjustment of these parameters.
    
For the {\hdiff} model (Sect. \ref{sec:hyper-model}), we originally used $\nu_1 = 0.03, \nu_2 = 0.2$, and $\eta_3 = 0.2$. 
The parameter $\nu_1$ affects the electrical resistivity as well as the viscous terms, and it scales up all the diffusive terms in the MHD equations over the entire computational domain. Therefore, this parameter should be kept as low as possible. It has been shown empirically that $\nu_1 > 0.02$ is needed to obtain stable solutions in several standard test problems to which Bifrost has been applied for a numerical solution \citep{2011A&A...531A.154G}. We studied different choices for this parameter for the 2D flux cancellation experiment and found that $\nu_1 = 0.03$ is a suitable choice because decreasing $\nu_1$ below this value leads to numerical instability in the current sheet, and increasing it much beyond this value will make the whole problem over-diffused.

Furthermore, it has been shown empirically that $\nu_2=0.2$ is about the minimum for numerically stable solutions in several standard test problems \citep{2011A&A...531A.154G}. In our case, the length of the current sheet is only slightly affected when this parameter was decreased below that value. However, running the experiment with a higher value of $\nu_2$ led to a reduction of the current sheet length, and therefore, to a considerable deviation between the numerical measures and analytical estimates shown in Figs.~\ref{fig:cs-analysis1} and \ref{fig:cs-analysis1-energy}.

The only free parameter of the {\hdiff} model that is interesting to adjust for our purposes is $\eta_3$ because it directly scales the electrical resistivity and has no effect on the viscosity. We tested running the experiment with different values of $\eta_3$ and obtained that values below 0.2 are numerically unstable, while values much higher than 0.2 increase the deviation between the numerical measures and the analytical estimates for the inflow values.

The simulation was also run using different values of $\eta_1$ for the {\syntelis} resistivity model. We found that this parameter can be decreased by an order of magnitude from the value used for the results in the above sections without losing numerical stability. However, this reduction of this diffusion parameter causes the current sheet length to be too long compared to the results of \citet{2019ApJ...872...32S}, thus deviating more from the analytically predicted current sheet length. A further decrease in $\eta_1$ will lead to numerical instability. When we instead increase this parameter by an order of magnitude, the current sheet length is too small compared to the analytical estimate. 
Decreasing the threshold value $J_{crit}$ has almost the same effect as increasing $\eta_1$.

The results obtained with the {\yokshi} resistivity model seem to be weakly dependent on
the scaling parameters: The current sheet length and Poynting influx barely increase when $\alpha$ is decreased by a factor ten. In addition, there is no significant change in the plasmoid behaviour. Decreasing this parameter further causes numerical instability. When the threshold value $v_{crit}$ is modified, it creates roughly the same effect as adjusting $\alpha$ the opposite way.

For each of the three resistivity models used in this experiment, we observed that the current sheet becomes numerically unstable when the anomalous resistivity is scaled down too strongly. This also shows that the experiment cannot be run without an anomalous resistivity for the given resolution because the current sheet would not be numerically resolvable, unless we were to use a uniform resistivity that is many orders of magnitude greater than the Spitzer resistivity, leading to very unphysical results, or if we were to increase the resolution by several orders of magnitude, causing the experiment to become expensive in terms of compute resources. 


\subsection{1D Harris current sheet}
\label{sec:Results-1Dharris}

\begin{figure*}[h]
    \centering
    \includegraphics[width=\textwidth]{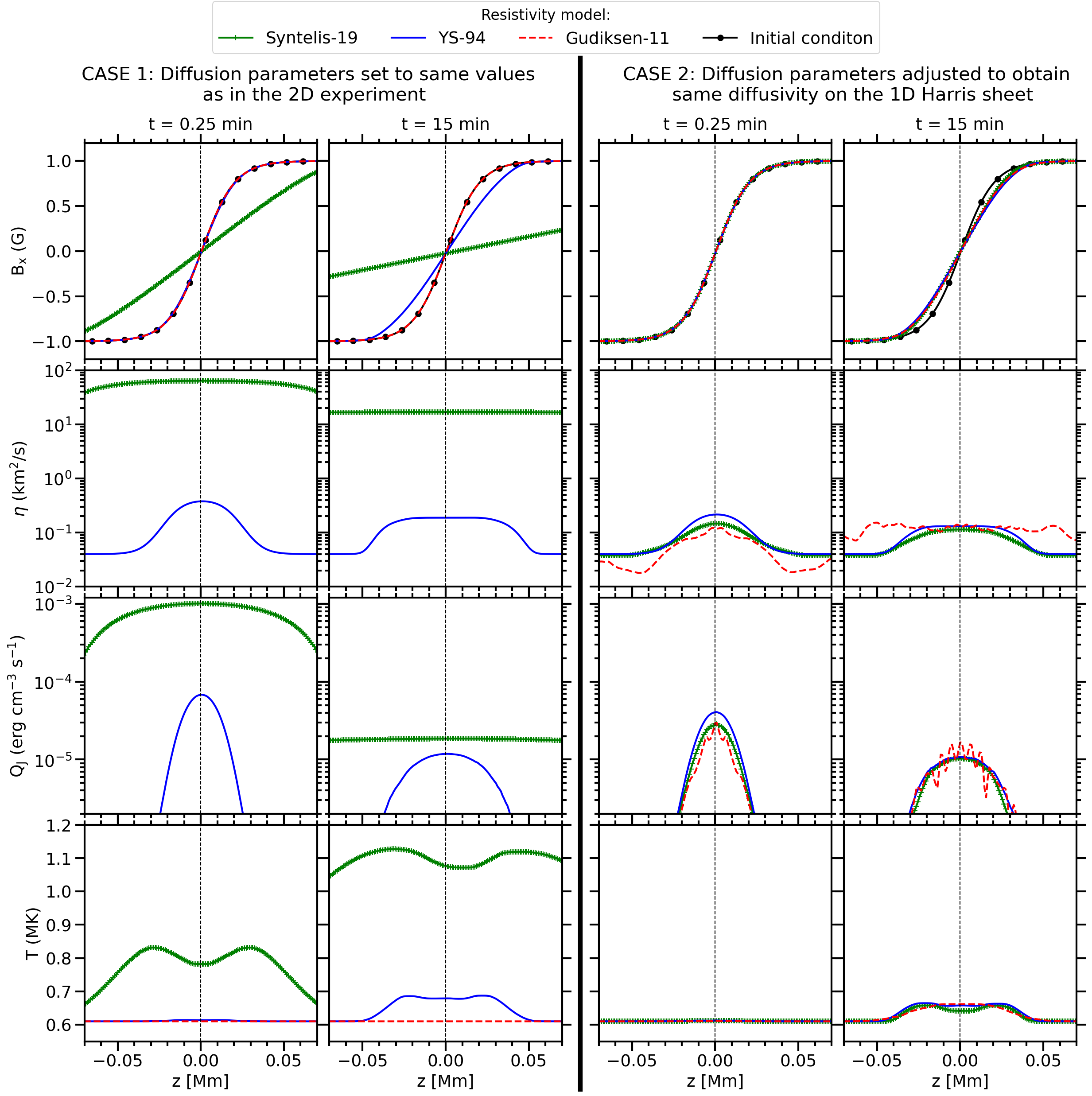}
    \caption{Evolution of the 1D Harris current sheet. From top to bottom, the magnetic field $B_x$, the resistivity $\eta$, the Joule heating $Q_J$, and the temperature $T$ are plotted as obtained by using the {\syntelis} (green), {\yokshi} (blue), and {\hdiff} (red) resistivity models. The first and second columns show the results, measured at different times, setting the diffusion parameters to the same values as used in the 2D experiment. The third and fourth columns show the results obtained after adjusting these diffusion parameters to obtain the same behaviour on this 1D Harris sheet for the three resistivity models.}
\label{fig:Harris_alpha_m=0k08_inflow}
\end{figure*}

In the previous section, we showed that we could use three different resistivity models in a 2D flux cancellation experiment and obtain relatively consistent results in terms of current sheet length and energy release by adjusting the diffusion parameters of each resistivity model. In this section, we begin to study the effects of applying the same resistivity models and parameters to the 1D Harris current sheet experiment introduced in Section \ref{sec:harris}. 

The results of the experiment for the magnetic field $B_x$, resistivity $\eta$, Joule heating $Q_J$, and temperature $T$ 
are shown in the first two columns of Fig.~\ref{fig:Harris_alpha_m=0k08_inflow} at two selected times: one time close to the beginning (0.25~min), and another time at the moment we stopped the simulation (15~min).
Even though we applied the same diffusion parameters that ensured relatively consistent results for the 2D flux cancellation experiment, the results for this 1D Harris sheet vary significantly depending on the resistivity model. At $t=0.25$ min, the {\syntelis} model has already had a huge impact in terms of diffusing out the current sheet width and heating up the plasma. The {\yokshi} model has a significant diffusive effect on the current sheet at $t=15$ min, but it is still small compared to the {\syntelis} model. The {\hdiff} model has apparently no diffusive effect on the current sheet with the given values for its free parameters. 
By fitting the $B_x$ profile to a hyperbolic tangent, $\tanh(z/w)$, and finding the width $w$ through the least-squares method, we find that the width of the current sheet, which initially is 20 km, 
has at $t=15$ min increased to 217 km with the {\syntelis} model and to 30 km with the {\yokshi} model, but it remained at 20 km with the {\hdiff} model.
The reason is that the resistivity (second row of the figure) in the {\syntelis} model is highest: it is up to two orders of magnitude higher than in the {\yokshi} model. At the end of the simulation, in the {\syntelis} case, its maximum is $\sim 15\ \mathrm{km^2\ s^{-1}}$, while for the {\yokshi} model, it is $\sim 0.20\ \mathrm{km^2\ s^{-1}}$. The resistivity stays $< 0.01\ \mathrm{km^2\ s^{-1}}$ in the {\hdiff} model. As a result of this,
the Joule heating, as seen in the third row, has a maximum value more than one order
of magnitude higher in the {\syntelis} case than in the {\yokshi} case at the early stages of the simulation, and then this difference decreases over time as the magnetic field is diffused and the currents are smaller. Since the resistivity is really low for the {\hdiff} case, the associated Joule heating in this case is negligible. Consequently, the temperature profile in the current sheet, which is initially uniform with a value of 0.61 MK, has risen to a maximum value above 1.1 MK in the {\syntelis} case at $t=15$ min, but only to 0.69 MK in the {\yokshi} case. It is unchanged in the {\hdiff} case.
The large asymmetry seen in the temperature profile for the {\syntelis} case at $t=15$ min is due to the tiny asymmetries in the staggered mesh, which are rapidly magnified by the relatively high diffusivity of this resistivity model (with the given values for the diffusion parameters).

For comparison, the third and fourth columns of  Fig.~\ref{fig:Harris_alpha_m=0k08_inflow} show the results after adjusting the scaling parameter of each resistivity model to ensure that they have roughly the same diffusive effect on this 1D Harris sheet. The new values for the adjusted parameters are $\eta_3 = 1.0$ for the {\hdiff} model, $\eta_1 = 3.78\times10^{-3}\ \mathrm{km^2\ s^{-1}}$ for the {\syntelis} model, and $\alpha = 2.0\times 10^{-8}\ \mathrm{km^2\ s^{-1}}$ for the {\yokshi} model. With the adjusted values, all three resistivity models diffuse the current sheet out to a final width of $\sim 26$ Mm at $t=15$ min.  The resistivity at the centre of the current sheet lies at slightly above $\sim 0.10\ \mathrm{km^2\ s^{-1}}$ in all three cases, causing the final Joule heating profiles to be nearly identical and the final maximum temperature to reach about 0.66 MK in all three cases.
One noticeable difference is seen in the resistivity in the regions outside the current sheet, where the magnetic field is nearly constant. The {\hdiff} model is nearly an order of magnitude higher than the other two models because the resistivity of this model depends, among other factors, on third derivatives of the magnetic field as well as on the gradients in the velocity perpendicular to the field. This makes it relatively sensitive to tiny perturbations in the current density that are enhanced by the velocity perturbations that arise during the diffusion of the current sheet. However, this  enhancement of the resistivity outside the current sheet does not affect the temperature profile at all because the current density, and hence the Joule heating, is here several orders of magnitude lower than at the centre of the current sheet. Additionally, the Lundquist number in the {\hdiff} case is above $10^4$ at any distance greater than 0.01 Mm away from the current sheet. This agrees well with the other two resistivity models. This shows indeed that the resistivity outside the current sheet has no effect on the evolution of the plasma.

We have shown that the resistivity models resulted in completely different
levels of the diffusive effect when they were applied in this 1D Harris current sheet experiment when the same diffusion parameter values were used that in the 2D flux cancellation experiment gave results that agreed well. We also demonstrated that we can easily adjust the diffusion parameters to obtain roughly the same diffusive behaviour in this relatively simple experiment. The free parameters of the {\yokshi} and {\hdiff} models only needed adjustments within roughly the same order of magnitude  to obtain these results, as shown in the second two columns of Fig.~\ref{fig:Harris_alpha_m=0k08_inflow}, but the $\eta_1$ value of the {\syntelis} model needed to be decreased by more than three orders of magnitude. This is due to its direct scaling with the current density, which causes the diffusivity of this resistivity model to be strongly dependent on the magnetic field topology.


\section{Discussion}
\label{sec:discussion}

This comparative study of resistivity models has demonstrated that we can use different types of resistivity models in the same numerical experiment and still obtain results that agree relatively well with each other. We successfully mimicked a 2D flux cancellation experiment from \citet{2019ApJ...872...32S} and found that using Bifrost's hyper-diffusive resistivity model \citep[referred to in this paper as {\hdiff}]{2011A&A...531A.154G} results in a current sheet length that more or less follows the same evolution as when using the current density-proportional resistivity model of the original experiment ({\syntelis}), given the right input values for the diffusion parameters. The magnetic field and velocity measured in the inflow region of the current sheet also develop in a similar way when the experiment is performed with each of these two resistivity models. As a result of this, the Poynting influx evolves similarly in both cases. The energy conversion, on the other hand, follows different schemes in each case. While the energy conversion in the {\syntelis} case agrees with the Petschek theory, the current sheet in the Gudiksen-11 case undergoes plasmoid-mediated reconnection and a significantly higher portion of the magnetic energy is converted into heat. As a result, the maximum temperature is higher in this last case.
The drift velocity-dependent resistivity model ({\yokshi}), previously applied by \citet{1994ApJ...436L.197Y}, among others, was also applied for the same experiment. The results obtained when using this resistivity model also agree satisfactorily with the results from the other two resistivity models. The current sheet is slightly shorter and the inflow magnetic field is slightly weaker, however, leading to a significantly lower Poynting influx. Despite undergoing plasmoid-mediated reconnection, a lower portion of the input magnetic energy is converted into heat in this case than in the Petschek-conform {\syntelis} case, in contrast to the {\hdiff} case, in which the conversion rate of magnetic energy to heat is higher. Therefore, the heated region has a lower temperature than in the other two cases. Except for the differences in terms of plasmoid generation and energy conversion, the temperature and mass density profiles of all three cases have a similar structural shape.

Furthermore, we observed that when we numerically solved the same model equations for a 1D Harris current sheet, the results in terms of diffusive rates and Joule heating obtained using each of the three resistivity models were significantly different from each other, given the same input values for the diffusion parameters as in the 2D experiment. Running the same experiment with adjusted values for the diffusion parameters showed that two of these resistivity models, namely {\hdiff} and {\yokshi}, needed only adjustments within the same order of magnitude for their scaling parameters in order to obtain the same diffusive rate on the Harris sheet. The  scaling parameter $\eta_1$ in the {\syntelis} resistivity model, on the other hand, needed to be scaled down by more than three orders of magnitude from its value applied in the 2D experiment in order to obtain the same diffusive rate in this 1D Harris sheet experiment as the other two resistivity models. 

One of the free parameters of the resistivity model used by \citet{2019ApJ...872...32S} requires an adjustment of several orders of magnitude when jumping between these two experiments because the resistivity scales linearly with the current density. This causes the ideal value for the scaling parameter to be strongly dependent of the magnetic field topology of the experiment when a satisfactory result is to be obtained. Moreover, its linear proportionality to the current density causes the resistivity to stay relatively high in relatively large areas around the current sheet. The Lundquist number therefore increases relatively slowly with distance from the current sheet compared to the other two resistivity models that were tested in this paper. Finally, because $\eta$ scales with the current density, the anomalous resistivity in regions near to magnetic sources needed to be turned off. This resistivity model works in a satisfactory way for several numerical experiments when the scaling parameter is adjusted properly, however.

We observed that the electron drift velocity-dependent resistivity model that was previously used by \citet{1994ApJ...436L.197Y} might be used to obtain results in both experiments of this paper that agree satisfactorily with the corresponding results obtained with Bifrost's hyper-diffusion model without adjusting the scaling parameter drastically. However, both our experiments dealt with coronal plasma with approximately the same temperature and density as well as similar magnetic field strength. The experiment of \citet{1994ApJ...436L.197Y}, on the other hand, which used the same resistivity model to handle current sheets in the upper convection zone, required the scaling parameter to be larger by several orders of magnitude. As the typical electron drift velocity and electron thermal velocity (which typically determines the threshold velocity at which this type of anomalous resistivity is to be activated)  differs by several orders of magnitude from the upper convection zone to the upper corona, the ideal values for the free parameters of this resistivity model strongly depend on the local plasma conditions. We were therefore also able to activate the anomalous resistivity of this model only in the coronal region of our 2D experiment (as the scaling parameter was set to handle coronal plasmas) and had to apply a relatively low uniform resistivity below. Despite this, we were fully able to use this resistivity model and obtain results in both our experiments that agreed relatively well with the results  obtained with the other two resistivity models, after the free parameters were adjusted properly.

The hyper-diffusive resistivity model of Bifrost \citep{2011A&A...531A.154G}, on the other hand, depends not only on the magnitude of magnetic field gradients, but also on the local fast-mode wave velocity, fluid velocity, and velocity gradients along magnetic field lines. This ensures that the resistivity of this model becomes large only when it is really needed to be large in order to make current sheets numerically resolvable and stay relatively low elsewhere.
With a default set of input values for the diffusion parameters, this resistivity model can be applied on anything from coronal plasmas to convection zone plasmas with any type of magnetic field topology without adjusting the parameters drastically. Therefore, this resistivity model does not need to be turned off and replaced by uniform resistivity in specific areas of the computational domain, but can rather be applied on the whole domain.

It is important to point out that several simplifications were made in this study, which is only a rough representation of driven reconnection in the solar atmosphere. For a more detailed study of the reconnection in the Sun, partially ionised effects such as ambipolar diffusion \citep{1989ApJ...340..550Z} and the Hall effect \citep{2011PhPl...18g2109H} cannot be ignored, especially when studying the energy balance in the chromosphere \citep{2023ApJ...946..115W} and the heating mechanisms for EBs \citep{2023RAA....23c5006L} and UV bursts \citep{2022A&A...665A.116N}. These effects also play a significant role in the structure of the inflow current density \citep{2018A&A...609A.100S}, plasmoid formation \citep{2019ApJ...884..161S, 2021PhPl...28c2901M}, and reconnection-driven slow-mode shocks \citep{2016A&A...591A.112H}. A detailed study of the reconnection rate in plasmoid-mediated reconnection may be performed with high-resolution simulations of a 2D current sheet \citep{2009AGUFMSM24B..07B}. More realistic studies of the turbulent energy cascade that occurs in flux ropes generated along the current sheets where the reconnection takes place can be made through high-resolution 3D MHD simulations \citep{2022SciA....8N7627D} or particle-in-cell simulations \citep{2011NatPh...7..539D}. We acknowledge that the details of the reconnection physics cannot be revealed through MHD models with anomalous resistivity, and this is not what we attempted to achieve with our study. With the simplifications and assumptions that were made, however, we achieved the insight that three relatively different anomalous resistivity models can be applied on a well-known physical problem to obtain results that agree relatively well with each other. The main gain in knowledge with the hyper-diffusive resistivity model of Bifrost from the results of our experiments is that it is not that strongly dependent on local plasma conditions and magnetic field topology and can therefore be applied on the whole solar atmosphere as well as to upper convection zone in numerical models without using different values for the free parameters in different areas of the computational domain.

%
%
\begin{acknowledgements}
This research has been supported by the European Research Council through the
Synergy Grant number 810218 (``The Whole Sun'', ERC-2018-SyG) and 
by the Research Council of Norway through its Centres of Excellence scheme, project
number 262622.
The simulations were performed on resources provided by  Sigma2 - the National Infrastructure for High Performance Computing and Data Storage in Norway.
The authors are grateful to the referee for his/her constructive comments to improve the paper.
\end{acknowledgements}

\bibliographystyle{aa}
\bibliography{Faerder_2023_paper1}

\end{document}